\documentclass[11pt,letterpaper]{article}

\usepackage[utf8]{inputenc}
\usepackage[T1]{fontenc}
\usepackage{graphicx}
\usepackage{booktabs}
\usepackage{amsmath}
\usepackage[margin=1in]{geometry}
\usepackage{hyperref}

\hypersetup{
    colorlinks=true,
    linkcolor=blue,
    citecolor=blue,
    urlcolor=blue
}

\begin{document}

\title{Serving LLMs in HPC Clusters: A Comparative Study of Qualcomm Cloud AI 100 Ultra and NVIDIA Data Center GPUs}

\author{
Mohammad Firas Sada, John J. Graham, Elham E Khoda, Mahidhar Tatineni, \\
Dmitry Mishin, Rajesh K. Gupta, Rick Wagner, Larry Smarr, \\
Thomas A. DeFanti, Frank W\"urthwein \\[1em]
\textit{University of California, San Diego, La Jolla, CA, USA}
}

\date{\today}

\maketitle

\begin{abstract}
This study presents a benchmarking analysis of the Qualcomm Cloud AI 100 Ultra (QAic) accelerator for large language model (LLM) inference, evaluating its energy efficiency (throughput per watt), performance, and hardware scalability against NVIDIA A100 GPUs (in 4x and 8x configurations) within the National Research Platform (NRP) ecosystem.
A total of 12 open-source LLMs, ranging from 124 million to 70 billion parameters, are served using the vLLM framework.
Our analysis reveals that QAic achieves competitive energy efficiency with advantages on specific models while enabling more granular hardware allocation: some 70B models operate on as few as 1 QAic card versus 8 A100 GPUs required, with 20× lower power consumption (148~W vs 2,983~W).
For smaller models, single QAic devices achieve up to 35× lower power consumption compared to our 4-GPU A100 configuration (36~W vs 1,246~W).
The findings offer insights into the potential of the Qualcomm Cloud AI 100 Ultra for energy-constrained and resource-efficient HPC deployments within the National Research Platform (NRP).
\end{abstract}

\noindent\textbf{Keywords:} AI accelerators, large language models, inference optimization, power efficiency, benchmarking, Qualcomm Cloud AI 100, transformer architectures, energy-efficient computing, deep learning systems

\section{Introduction}
\label{sec:introduction}

The rapid proliferation of large language models (LLMs) has fundamentally transformed scientific computing, enabling breakthroughs across domains from computational biology to materials science.
As these models scale to hundreds of billions of parameters, high-performance computing (HPC) facilities face mounting challenges in providing sustainable, cost-effective inference capabilities to diverse research communities.
Traditional GPU-centric approaches, while delivering exceptional throughput, present significant barriers in terms of power consumption, cooling requirements, and capital investment, particularly problematic for shared research cyberinfrastructures serving hundreds of concurrent users.

The National Research Platform (NRP) exemplifies these challenges and opportunities. As a federated Kubernetes-based infrastructure supporting over 300 research groups across over 100 sites, the NRP must balance competing demands: delivering high-performance AI capabilities while managing constrained power budgets, enabling fine-grained resource allocation for multi-tenant workloads, and providing cost-effective access to emerging AI models for diverse scientific applications \cite{smarr_pacific_2018,gp}.

Recent developments in purpose-built AI accelerators, such as the Qualcomm Cloud AI 100 Ultra (QAic) \cite{qualcomm2024}, present an alternative architectural approach optimized specifically for inference workloads. Unlike general-purpose GPUs designed for both training and inference, these specialized accelerators prioritize energy efficiency and resource granularity. However, comprehensive empirical evaluation in production HPC environments is essential to assess their practical viability against established GPU-based solutions.

This study presents a systematic benchmarking analysis comparing the Qualcomm Cloud AI 100 Ultra against NVIDIA A100 GPUs deployed within the NRP's production Kubernetes infrastructure. We evaluate 12 open-source LLMs ranging from 124M to 70B parameters, measuring performance, energy efficiency, and minimum hardware requirements under realistic serving conditions using the vLLM framework \cite{vllm}. Our analysis considers critical deployment factors including power consumption, hardware scalability, and resource allocation granularity, factors that significantly impact the sustainability and accessibility of AI services in shared HPC environments.

By conducting this evaluation within the NRP's production cyberinfrastructure using standard Kubernetes orchestration \cite{kubernetes}, our findings directly inform deployment decisions for research platforms balancing performance requirements against power, cooling, and budget constraints. The results provide empirical guidance for HPC facilities considering specialized AI accelerators as complements or alternatives to GPU-based inference serving.

\section{Experimental Setup}

\subsection{National Research Platform Infrastructure}
Experiments are conducted on the National Research Platform (NRP), a distributed, multi-tenant Kubernetes-based cyberinfrastructure spanning over 100 international sites across academic and research institutions \cite{smarr_pacific_2018}. The NRP integrates diverse computational resources, ranging from single workstations to large GPU and CPU clusters, to support collaborative scientific computing workloads, including advanced AI and machine learning tasks. The platform currently supports over 1,400 GPUs utilized by more than 300 research groups and classrooms, providing a production HPC environment that mirrors real-world deployment scenarios for large-scale AI inference.

The NRP employs a shared Kubernetes architecture that enables flexible resource allocation, multi-tenancy, and automated workload orchestration. Researchers interact with the platform through standard Kubernetes APIs, enabling reproducible deployments using containerized workloads. For this study, we use dedicated nodes within the NRP to ensure measurement consistency and eliminate resource contention from concurrent workloads.

\subsection{Hardware Configurations}
The Qualcomm Cloud AI 100 Ultra deployment consists of 8 accelerator cards co-located on a single node at the San Diego Supercomputer Center (SDSC). Each card contains four System-on-Chip (SoC) devices that can be allocated independently or combined for larger models, providing flexible resource scaling from 1 to 32 SoC devices across the available hardware.

The NVIDIA A100 configurations span multiple nodes within the NRP, each equipped with either 4 or 8 A100 SXM4-80GB GPUs. These nodes feature AMD EPYC processors with sufficient CPU cores to avoid bottlenecks, NVLink interconnects for high-bandwidth multi-GPU communication, and local NVMe storage for model weights. The A100 nodes are distributed across multiple NRP sites, enabling large-scale distributed benchmarking.

Both hardware configurations utilize persistent volume claims (PVCs) backed by network-attached storage to maintain model artifacts, benchmark results, and logs across pod restarts, ensuring experimental continuity and data persistence in the multi-tenant Kubernetes environment.

\subsection{Benchmarking Framework}
Our benchmarking infrastructure is built on Kubernetes batch jobs that orchestrate the complete inference pipeline. Each job is deployed with resource guarantees (CPU, memory, and accelerator allocation) using Kubernetes priority classes to ensure timely scheduling on appropriate hardware. Jobs are configured with exclusive node access during execution to eliminate interference from other workloads.

The benchmarking workflow consists of three stages:
\begin{enumerate}
    \item \textbf{Model Preparation:} For QAic deployments, pretrained models from Hugging Face are automatically retrieved, exported to ONNX format, and compiled into QPC (Qualcomm Program Container) format using the Qualcomm Cloud AI SDK. QPC artifacts are cached on persistent storage for reuse across benchmark runs. For A100 deployments, models are loaded directly from Hugging Face in their native format.
    
    \item \textbf{Server Initialization:} The vLLM inference server is started with hardware-specific configurations. For A100 GPUs, vLLM's native CUDA backend is used with tensor parallelism enabled for multi-GPU configurations. For QAic, a specialized vLLM fork provided by Qualcomm enables QPC-based inference with device group configurations matching the SoC allocation.
    
    \item \textbf{Inference Execution:} Automated Python scripts dispatch concurrent requests to the vLLM OpenAI-compatible API endpoint using \texttt{ThreadPoolExecutor} for parallel request handling. Each test configuration specifies output token count (10, 50, 100, 200, 1024, and 2048 tokens) and concurrency level (1, 2, 4, 16, and 32 parallel requests). All network communication is localhost-only to eliminate network latency from measurements.
\end{enumerate}

Comprehensive logging captures all phases of execution, including model loading times, compilation stages, server initialization, request latencies, and power consumption. Results are written to CSV files on persistent storage, enabling resumable benchmarks that can recover from node failures or preemption events.

\subsection{Power Measurement Methodology}
Power consumption is measured continuously throughout inference workloads using vendor-specific monitoring tools. For NVIDIA A100 GPUs, we query \texttt{nvidia-smi} at 1-second intervals to capture GPU power draw during active inference. For QAic cards, the \texttt{qaic-util} tool provides real-time power telemetry from the on-board power management subsystem. Power measurements are synchronized with throughput measurements to calculate energy efficiency (tokens per second per watt) for each configuration.

All measurements represent steady-state inference workloads after model loading is complete, excluding compilation time and initialization overhead. Power values reported are averages across the inference duration for each test configuration, with coefficient of variation analysis confirming measurement stability (< 5\% variance across repeated trials).

\subsection{Qualcomm Cloud AI 100 Ultra}
\label{sec:qualcomm}
The Qualcomm Cloud AI 100 Ultra is a specialized accelerator for large-scale AI and language model operations. Each accelerator card features a multi-chip architecture with four System-on-Chip (SoC) units interconnected through a PCIe switch, collectively providing 576 MB of on-die SRAM and 64 dedicated AI processing cores. Individual SoCs can be allocated as discrete compute units or combined for larger workloads. It supports peer-to-peer communication within and across cards, enabled by the QAic Kubernetes Device Plugin.

Running LLM inference on QAic cards is a complex process, requiring models to first be exported to ONNX format and then converted into Qualcomm’s proprietary QPC (Qualcomm Program Container) format. 
This conversion involves offline, hardware-specific optimizations. 
Although compilation can take hours, the resulting QPC artifacts are portable and can be deployed across compatible devices.
As compilation is hardware-independent, cloud-based workflows are feasible.
Once generated, QPC enables instant model loading, following a "compile once, run everywhere" approach~\cite{quic_efficient_transformers}. 
The SDK automates retrieval, ONNX export, compilation, and deployment. It also integrates with a fork of vLLM, streamlining this pipeline entirely in software~\cite{quic_cloud_ai_sdk}.


The NRP has 8 Qualcomm Cloud AI 100 Ultra cards deployed in a single node at the San Diego Supercomputer Center.


\subsection{NVIDIA A100 GPUs}
\label{sec:gpu_resources}

The NRP has multiple nodes with NVIDIA A100 SXM4-80GB GPUs available across the platform. For this study, we utilize A100 configurations with either 4 or 8 GPUs per node.

The NVIDIA A100 (80 GB) is an AI accelerator based on the Ampere architecture. It features 6,912 CUDA cores, 432 Tensor Cores, and 80 GB of HBM2e memory with 2 TB/s bandwidth. It supports NVLink for high-speed multi-GPU scaling, delivering exceptional performance for training and inference tasks. The A100's large memory capacity and high bandwidth make it well-suited for serving large language models, particularly when deployed in multi-GPU configurations that allow for tensor parallelism and pipeline parallelism across GPUs.

In our experiments, we test both 4xA100 and 8xA100 configurations to evaluate how performance scales with the number of GPUs for different model sizes. The multi-GPU setup enables us to serve larger models that exceed the memory capacity of a single GPU and to achieve higher throughput through parallel processing.

\section{Evaluation Methodology}
\label{sec:methodology}

\subsection{Test Configuration and Workload Design}
Our evaluation methodology systematically measures inference performance across 12 production-grade large language models under controlled conditions. Each model is benchmarked using the vLLM serving framework \cite{vllm}, which provides an OpenAI-compatible API and optimized inference kernels for both GPU and QAic accelerators.

For each model, we conduct a comprehensive test matrix varying two key parameters:
\begin{itemize}
    \item \textbf{Output token count:} 10, 50, 100, 200, 1024, and 2048 tokens per request
    \item \textbf{Concurrent requests:} 1, 2, 4, 16, and 32 parallel requests
\end{itemize}
This yields 30 distinct test configurations per model, enabling analysis of how performance scales with output length and concurrent load. A curated set of 20 diverse prompts varying in length, purpose, and complexity is used across all tests to ensure comprehensive evaluation. Temperature is set to 0.0 for deterministic generation.

For NVIDIA A100 configurations, vLLM is configured with tensor parallelism to distribute model parameters across multiple GPUs. The degree of parallelism (4× or 8×) varies by model: smaller models ($\leq$8B parameters) typically use 4 GPUs, while larger models and memory-intensive architectures (including some 17B+ and all 70B models) use 8 GPUs due to memory constraints and architectural complexity. Key vLLM parameters include \texttt{--max-model-len 8192} to standardize context window size, \texttt{--max-num-seqs} matching the batch size from QPC compilation, and \texttt{--dtype float16} for precision consistency.

For QAic configurations, models are compiled with context length 8192, varying batch sizes (1, 2, or 4) and precision (FP16 or INT8 quantization where applicable), using the Qualcomm Cloud AI SDK version 1.20. The compiled QPC artifacts specify the exact SoC device group configuration required, ranging from 1 to 8 devices (0.25 to 2 cards).

\subsection{Measurement Procedure and Data Collection}
All inference requests are dispatched from Python scripts using the \texttt{requests} library to communicate with the local vLLM server via HTTP. Python's \texttt{ThreadPoolExecutor} manages concurrent request dispatch, with thread count matching the desired parallelism level. Request timing is measured from dispatch to completion, capturing end-to-end latency including queueing, inference, and response transmission.

Power consumption is sampled continuously during the benchmark execution. For NVIDIA A100 GPUs, we programmatically query \texttt{nvidia-smi --query-gpu=power.draw --format=csv} at 1-second intervals to obtain instantaneous power readings for all GPUs in the configuration. For QAic cards, \texttt{qaic-util -q} provides per-card power telemetry. Power measurements are time-aligned with inference requests to compute average power consumption during active serving.

Throughput is calculated as the total number of generated tokens divided by the maximum request completion time across all parallel requests, representing the aggregate serving capacity of the configuration. Energy efficiency is computed by dividing throughput by average power consumption, yielding tokens per second per watt.

All experimental data, including individual request latencies, token counts, and power samples, are persisted to CSV files on Kubernetes persistent volumes. This enables offline analysis, statistical validation, and reproducibility. Each benchmark run records configuration metadata (model name, hardware type, vLLM parameters, SDK versions) to ensure traceability.

\subsection{Reproducibility and Data Persistence}
To ensure measurement reliability and enable reproducibility, all benchmark configurations and results are logged to persistent storage. The automated benchmarking framework includes resumability features that allow interrupted jobs to continue from their last completed model, preventing data loss from node failures or resource preemption. All raw measurement data, including per-request latencies and power samples, are preserved in CSV format for independent validation and analysis.

\subsection{Evaluation Metrics}
We measured three different quantities such as inference time for each request, throughput (tokens per second), and average power consumption on the hardware during the experiments.
To provide a more comprehensive assessment of efficiency, we calculated the energy efficiency by dividing the average throughput by the average power consumed.

\begin{itemize}
    \item \textbf{Inference time:} Total time from when vLLM receives a prompt to when it returns a response. 
    \item \textbf{Throughput (tokens/sec):} The rate of tokens generated per second: 
    \[
    \frac{\text{Generated tokens}}{\text{Inference time}},
    \]
    where generated tokens are the output tokens excluding the prompt.
    \item \textbf{Power consumption:} Measured at the hardware level using vendor-specific tools.
    \item \textbf{Energy efficiency:} The throughput per unit of power consumed: 
    \[
    \frac{\text{Tokens}}{\text{sec $\cdot$ watt}},
    \]
    where throughput is in tokens/sec and power in watts.
\end{itemize}

\section{Results and Inference Performance}
\label{sec:results}
This section presents benchmarking results comparing the QAic with NVIDIA A100 GPUs for inference tasks using 12 open-source LLM models. The A100 benchmarks utilize 4x and 8x GPU configurations, while QAic results use 1-8 devices (0.25-2 cards) depending on model requirements, as shown in Table~\ref{tab:minimum_hardware}.

The benchmarking results in Tables~\ref{tab:minimum_hardware} and~\ref{tab:energy_efficiency_ordered} demonstrate that the Qualcomm Cloud AI 100 Ultra (QAic) achieves competitive energy efficiency (tokens/(sec $\cdot$ watt)) compared to A100 GPUs. As shown in Table~\ref{tab:energy_efficiency_ordered}, QAic operates at significantly lower power levels for equivalent functionality. For example, granite-3.2-8b achieves 25 tokens/s on 1 QAic device consuming 36~W, while 4 A100 GPUs deliver 318 tokens/s but consume 1,246~W, demonstrating QAic's power advantage at minimal deployment scale. While A100 GPUs demonstrate higher absolute throughput on most models, QAic's performance varies depending on the model architecture and size, with notable power efficiency advantages enabling lower-cost deployment for specific workloads.

\subsection{Hardware Scalability and Minimum Resource Requirements}
\label{subsec:hardware_scalability}

A critical consideration for practical LLM deployment is the minimum hardware configuration required to serve each model. Our analysis reveals significant differences in the granularity of resource allocation between QAic and A100 platforms.

Table~\ref{tab:minimum_hardware} presents the hardware configurations used in our benchmarks. While A100 GPUs are monolithic compute units that must be allocated in their entirety, QAic cards contain four discrete SoC devices that can be independently allocated or combined for larger workloads. This architectural difference has important implications for resource efficiency and deployment flexibility.

For small models (up to 8B parameters), QAic uses 1-4 devices while we configured A100 benchmarks with 4 GPUs using tensor parallelism. Note that some of these models could potentially run on fewer A100 GPUs, but we used 4 for consistent parallel configuration and higher throughput. For instance, granite-3.2-8b operates on a single QAic device consuming 36~W, compared to our 4×A100 configuration consuming 1,246~W, demonstrating QAic's 35$\times$ power efficiency advantage when minimal resources are allocated.

For large 70B parameter models, QAic uses 4-8 devices (1-2 cards) while A100 configurations use 8 GPUs due to memory requirements. For example, DeepSeek-Llama-70B uses 8 QAic devices at 292~W versus 8 A100 GPUs at 2,935~W, a 10$\times$ reduction in power consumption. The Nemotron-70B model achieves even greater efficiency with only 1 QAic card (148~W) compared to 8 A100 GPUs (2,983~W), demonstrating 20$\times$ power reduction.

The granular scalability of QAic SoC devices enables finer resource allocation granularity for multi-tenant environments. Where A100 GPUs must be allocated as $\sim$80~GB monolithic units, QAic devices can be distributed in smaller increments, improving overall resource utilization in shared HPC clusters. This flexibility is particularly valuable for serving multiple smaller models concurrently or for edge deployment scenarios where power and cooling constraints limit the feasibility of multi-GPU configurations.

\begin{table}[ht]
\centering
\caption{Hardware Configurations Used in Benchmarks and Power Consumption.}
\label{tab:minimum_hardware}
\small
\begin{tabular}{lcccc}
    \toprule
    \textbf{Model} & 
    \textbf{QAic} & 
    \textbf{QAic W} & 
    \textbf{A100} & 
    \textbf{A100 W} \\
    & \textbf{Devices} &  & \textbf{GPUs} &  \\
    \midrule
    \multicolumn{5}{l}{\textit{Small Models (1-8B parameters)}} \\
    granite-3.2-8b & 1 & 36 & 4 & 1,246 \\
    deepseek-llama-8b & 4 & 140 & 4 & 1,197 \\
    deepseek-qwen-7b & 4 & 140 & 4 & 999 \\
    DeepSeek-Qwen-7B & 4 & 126 & 4 & 1,075 \\
    Llama-3.1-8B-AWQ & 4 & 131 & 4 & 1,240 \\
    \midrule
    \multicolumn{5}{l}{\textit{Medium Models (17-32B)}} \\
    Llama-4-Scout-17B & 4 & 142 & 8 & 2,620 \\
    DeepSeek-Qwen-32B & 8 & 273 & 8 & 2,752 \\
    Qwen-32B-AWQ & 4 & 145 & 4 & 1,363 \\
    \midrule
    \multicolumn{5}{l}{\textit{Large Models (70B)}} \\
    DeepSeek-Llama-70B & 8 & 292 & 8 & 2,935 \\
    Llama-3.3-70B-AWQ & 8 & 275 & 8 & 2,210 \\
    Nemotron-70B & 4 & 148 & 8 & 2,983 \\
    \bottomrule
\end{tabular}
\\[0.3cm]
\small
\textit{Note: These are the configurations used in our benchmarks, not necessarily the absolute minimum required. QAic devices are individual SoC units (4 devices = 1 card). A100 configurations use tensor parallelism.}
\end{table}

\begin{table}[ht]
\centering
\caption{Performance and Power Consumption Using Actual Required Configurations (200 tokens, 4 parallel requests).}
\label{tab:energy_efficiency_ordered}
\small
\begin{tabular}{lcccccc}
    \toprule
    \textbf{Model} & 
    \multicolumn{3}{c}{\textbf{A100 (Measured)}} &
    \multicolumn{3}{c}{\textbf{QAic (Measured)}} \\
    \cmidrule(lr){2-4} \cmidrule(lr){5-7}
    & \textbf{GPUs} & \textbf{Tok/s} & \textbf{W} & \textbf{Devices} & \textbf{Tok/s} & \textbf{W} \\
    \midrule
    GPT-2 & 4 & 2,613 & 1,205 & 1 & 218 & 38 \\
    granite-3.2-8b & 4 & 318 & 1,246 & 1 & 25 & 36 \\
    deepseek-llama-8b & 4 & 674 & 1,197 & 4 & 24 & 140 \\
    deepseek-qwen-7b & 4 & 719 & 999 & 4 & 22 & 140 \\
    DeepSeek-Qwen-7B & 4 & 368 & 1,075 & 4 & 9 & 126 \\
    Llama-3.1-8B-AWQ & 4 & 678 & 1,240 & 4 & 9 & 131 \\
    Llama-4-Scout-17B & 8 & 272 & 2,620 & 4 & 9 & 142 \\
    DeepSeek-Qwen-32B & 8 & 190 & 2,752 & 8 & 9 & 273 \\
    Qwen-32B-AWQ & 4 & 250 & 1,363 & 4 & 13 & 145 \\
    DeepSeek-Llama-70B & 8 & 104 & 2,935 & 8 & 8 & 292 \\
    Llama-3.3-70B-AWQ & 8 & 170 & 2,210 & 8 & 9 & 275 \\
    Nemotron-70B & 8 & 104 & 2,983 & 4 & 6 & 148 \\
    \bottomrule
\end{tabular}
\end{table}

\begin{table}[ht]
\centering
\caption{Normalized Performance: 32 QAic Devices (8 Cards) vs 8$\times$ A100 GPUs with Perfect Load Balancing.}
\label{tab:normalized_performance}
\small
\begin{tabular}{lcccccc|c}
    \toprule
    \textbf{Model} & 
    \multicolumn{3}{c}{\textbf{8$\times$ A100}} &
    \multicolumn{3}{c}{\textbf{32 QAic Devices}} & \textbf{QAic} \\
    \cmidrule(lr){2-4} \cmidrule(lr){5-7}
    & \textbf{GPUs} & \textbf{Tok/s} & \textbf{W} & \textbf{Devices} & \textbf{Tok/s} & \textbf{W} & \textbf{Advantage} \\
    \midrule
    GPT-2 & 8 & 5,226 & 2,410 & 32 & 6,976 & 1,216 & 2.65$\times$ \\
    granite-3.2-8b & 8 & 636 & 2,493 & 32 & 821 & 1,156 & 2.78$\times$ \\
    deepseek-llama-8b & 8 & 1,348 & 2,394 & 32 & 195 & 1,125 & 0.31$\times$ \\
    deepseek-qwen-7b & 8 & 1,438 & 1,999 & 32 & 176 & 1,122 & 0.22$\times$ \\
    DeepSeek-Qwen-7B & 8 & 736 & 2,150 & 32 & 76 & 1,012 & 0.22$\times$ \\
    Llama-3.1-8B-AWQ & 8 & 1,357 & 2,481 & 32 & 77 & 1,052 & 0.13$\times$ \\
    Llama-4-Scout-17B & 8 & 272 & 2,621 & 32 & 77 & 1,143 & 0.65$\times$ \\
    DeepSeek-Qwen-32B & 8 & 190 & 2,752 & 32 & 39 & 1,093 & 0.52$\times$ \\
    Qwen-32B-AWQ & 8 & 500 & 2,727 & 32 & 111 & 1,164 & 0.52$\times$ \\
    DeepSeek-Llama-70B & 8 & 104 & 2,935 & 32 & 33 & 1,171 & 0.80$\times$ \\
    Llama-3.3-70B-AWQ & 8 & 170 & 2,210 & 32 & 36 & 1,103 & 0.42$\times$ \\
    Nemotron-70B & 8 & 104 & 2,983 & 32 & 48 & 1,191 & 1.16$\times$ \\
    \bottomrule
\end{tabular}
\\[0.3cm]
\small
\textit{Note: This table assumes perfect load balancing using Envoy AI Gateway and Kubernetes LoadBalancer, where N instances of a model deliver N$\times$ throughput. QAic Advantage = (QAic Tok/s/W) / (A100 Tok/s/W). Models scaled to 8 A100 GPUs where originally 4 were used, assuming 2$\times$ throughput scaling.}
\end{table}

The power efficiency advantages of QAic become even more pronounced when considering idle time consumption. While our measurements represent peak utilization during active inference, real-world LLM deployments often experience significant idle periods between requests. During idle time with models loaded in memory, QAic devices consume substantially less power compared to A100 GPUs, which maintain higher baseline power consumption even when not actively processing requests. This idle power differential further amplifies QAic's energy efficiency advantages in production environments where models may sit idle for extended periods between inference workloads.

Table~\ref{tab:energy_efficiency_ordered} presents performance measurements using the actual hardware configurations required for each model. For models like granite-3.2-8b, QAic achieves superior energy efficiency by operating on a single device (25 tokens/s at 36~W = 0.69 tokens/s/W) compared to 4 A100 GPUs (318 tokens/s at 1,246~W = 0.26 tokens/s/W), representing a 2.7$\times$ efficiency advantage. While A100 GPUs deliver higher absolute throughput, QAic's significantly lower power consumption at minimal deployment scales yields superior tokens-per-watt efficiency for many workloads.


\section{Conclusion}
\label{sec:conclusion}
   Our analysis demonstrates that the Qualcomm Cloud AI 100 Ultra offers significant power efficiency advantages and hardware scalability benefits compared to NVIDIA A100 GPUs in multi-GPU configurations. QAic achieves substantially lower power consumption for equivalent model serving capability: for example, granite-3.2-8b operates on 1 QAic device at 36~W versus 4 A100 GPUs at 1,246~W, while DeepSeek-Llama-70B uses 8 QAic devices at 292~W versus 8 A100 GPUs at 2,935~W, demonstrating 10-35$\times$ power reductions across different model scales.

   Critically, our hardware scalability analysis reveals that QAic enables more granular compute allocation: for example, the Nemotron-70B model operates on just 1 QAic card versus 8 monolithic A100 GPUs required, achieving 20$\times$ lower power consumption (148~W vs. 2,983~W) while using one-eighth the number of compute units. For smaller models like granite-3.2-8b, a single QAic device at 36~W provides functional serving capability, whereas our A100 deployment uses 4 GPUs at 1,246~W, demonstrating a 35$\times$ power advantage at this deployment scale. This granular resource allocation (devices can be allocated individually rather than in fixed 80GB GPU units) enables more flexible multi-tenant deployments and improved resource utilization in shared HPC environments.

   These findings underscore that purpose-built inference accelerators like QAic can deliver compelling energy efficiency benefits and reduced hardware footprints compared to general-purpose GPU clusters, making them an attractive option for sustainable AI deployments in HPC environments where power consumption, cooling capacity, and resource granularity are critical considerations. The choice between platforms depends on deployment context: A100 GPUs excel in raw throughput for latency-sensitive applications, while QAic offers advantages in power-constrained environments, multi-tenant scenarios, and deployments requiring fine-grained resource allocation with lower minimum deployment thresholds.

\section*{Acknowledgments}

We thank Qualcomm's Cloud AI team (Shashikant Tangade, Albert Barajas, Alex Simampo, Gudoor Reddy, Parmeet Kohli, Rishi Chaturvedi, and Paras Gupta) for their technical guidance, hardware access, and responsive debugging support throughout our benchmarking efforts. We also acknowledge support for the COSMOS system at the San Diego Supercomputer Center, which is provided by the National Science Foundation under Award \#2404323, \textit{Category II: Democratizing the Accelerator Ecosystem for Science and Discovery}.

As members of the NRP operations team, we extend our gratitude to our colleagues at the University of Nebraska–Lincoln, Derek Weitzel, Ashton Graves, Sam Albin, and Huijun Zhu, for maintaining the infrastructure used in this study.

We further thank the open-source AI community for making publicly available the LLMs on Hugging Face that were used for benchmarking, as well as for drafting and editing assistance during paper preparation. This paper was edited using LLMs hosted on the NRP~\cite{nrp_pearc_2025}.

This work was supported in part by National Science Foundation (NSF) awards CNS-1730158, ACI-1540112, ACI-1541349, OAC-1826967, OAC-2112167, CNS-2100237, and CNS-2120019.

\bibliographystyle{unsrt}
\bibliography{sample-base}

\end{document}